\documentclass[conference]{IEEEtran}
\IEEEoverridecommandlockouts

\usepackage[utf8]{inputenc}
\usepackage{nicefrac}
\usepackage{comment}
\usepackage{makecell}

\usepackage{cite}
\usepackage{amsmath,amssymb,amsfonts,amsthm}
\usepackage{algorithmic}
\usepackage{graphicx}
\usepackage{textcomp}
\usepackage{xcolor}
\def\BibTeX{{\rm B\kern-.05em{\sc i\kern-.025em b}\kern-.08em
    T\kern-.1667em\lower.7ex\hbox{E}\kern-.125emX}}

\newcommand\In {:}
\newcommand\Wide[1] {~~#1~~}
\newcommand\Defs {:=}

\newcommand\Apply {\triangleright}

\newcommand\Dist {\mathbb{D}}

\newcommand\Hyper[2]  {[#1{\Apply}#2]}
\newcommand\Leak {\mathcal{L}}
\newcommand\Lift {\mathrm{Lift}}

\newcommand\mult    {\times}
\newcommand\MAX    {\mathrm{max}}

\newcommand\MaxLeak {\Leak^\MAX}
\newcommand\MaxLift {\mathcal{M}\Lift}
\newcommand\RealNN {\mathbb{R}_{\geq 0}}

\newcommand\VMax  {V^\MAX}
\newcommand\VgMax  {\VMax_g}

\newcommand\Sec[1] {\S\ref{#1}}
\newcommand\Eqn[1] {Eqn~(\ref{#1})}
\newcommand\Thm[1] {Thm~\ref{#1}}
\newcommand\Lem[1] {Lem~\ref{#1}}
\newcommand\Def[1] {Defn~\ref{#1}}
\newcommand\Cor[1] {Cor~\ref{#1}}

\newcommand\calx {\mathcal{X}}
\newcommand\caly {\mathcal{Y}}
\newcommand\calz {\mathcal{Z}}

\newcommand\calw {\mathcal{W}}

\newif\ifqif

\qiftrue

\ifqif
    \newcommand\px      {\pi_x}    
    \newcommand\py      {p(y)}
    \newcommand\CondYX  {C_{x,y}}
    
    \newcommand\MultBayes {\Bayes}
\else
    \newcommand\px   {P_X(x)}
    \newcommand\py   {P_Y(y)}
    
    \newcommand\CondYX  {P_{Y|X}(y|x)}
    \newcommand\MultBayes {I^{\mathrm{Sibson}}{\infty}}
\fi

\newenvironment{Reason}{\begin{tabbing}\hspace{2em}\= \hspace{1cm} \= \kill}
{\end{tabbing}\vspace{-1em}}
\newcommand\Step[2] {#1 \> $\begin{array}[t]{@{}llll}#2\end{array}$ \\}
\newcommand\StepR[3] {#1 \> $\begin{array}[t]{@{}llll}#3\end{array}$
\` {\RF \makebox[0pt][r]{\begin{tabular}[t]{r}``#2''\end{tabular}}} \\}
\newcommand\WideStepR[3] {#1 \>
$\begin{array}[t]{@{}ll}~\\#3\end{array}$ \`
{\RF \makebox[0pt][r]{\begin{tabular}[t]{r}``#2''\end{tabular}}} \\}

\newcommand\RF {\small}

\DeclareMathOperator*{\argmin}{arg\,min}

\newtheorem{definition}{Definition}
\newtheorem{theorem}{Theorem}
\newtheorem{lemma}{Lemma}
\newtheorem{corollary}{Corollary}
\newtheorem{example}{Example}

\newcommand\App[1]  {Appendix~\ref{#1}}

\begin{document}

\title{Explaining $\epsilon$ in local differential privacy through the lens of quantitative information flow}

\author{\IEEEauthorblockN{Natasha Fernandes}
\IEEEauthorblockA{\textit{School of Computing} \\
\textit{Macquarie University}\\
Australia \\
natasha.fernandes@mq.edu.au}
\and
\IEEEauthorblockN{Annabelle McIver}
\IEEEauthorblockA{\textit{School of Computing} \\
\textit{Macquarie University}\\
Australia \\
annabelle.mciver@mq.edu.au}
\and
\IEEEauthorblockN{Parastoo Sadeghi}
\IEEEauthorblockA{\textit{School of Engineering and IT} \\
\textit{UNSW Canberra}\\
Australia \\
p.sadeghi@unsw.edu.au}
}

\maketitle
\pagestyle{empty}
\thispagestyle{empty}
%
\begin{abstract}

The study of leakage measures for privacy has been a subject of intensive research and is an important aspect of understanding how privacy leaks occur in computer systems. Differential privacy has been a focal point in the privacy community for some years and yet its leakage characteristics are not completely understood. In this paper we bring together two areas of research --information theory and the $g$-leakage framework of  quantitative information flow (QIF)-- to give an operational interpretation for the epsilon parameter of local differential privacy. We find that epsilon emerges as a capacity measure in both frameworks; via (log)-lift, a popular measure in information theory; and via max-case $g$-leakage, which we introduce to describe the  leakage of any system to Bayesian adversaries modelled using ``worst-case'' assumptions under the QIF framework. 
Our characterisation resolves an important question of interpretability of epsilon and consolidates a number of disparate results covering the literature of both information theory and quantitative information flow. 


\end{abstract}

\section{Introduction}\label{s1827}

Over the past two decades, characterising and limiting privacy leakage in data sharing systems has been the subject of intensive research. Information-theoretic privacy \cite{2012_privacy_statisticalinference,2014PFInfBottneck}, differential privacy \cite{Dwork06} and quantitative information flow (QIF) \cite{Smith09} are three main frameworks that have been developed and studied for this purpose. 

Information-theoretic privacy is concerned with measuring privacy leakage using well-known information-theoretic quantities, such as Shannon mutual information \cite{thomas2006elements}:
\[I(X;Y) \Wide\Defs \sum_{x\in \calx, y \in \caly~}p(x,y)\log\frac{p(x,y)}{p(x)p(y)}~,\]
where $X$ denotes a secret and $Y$ denotes the observable, with $p(x,y), p(x)$ and $p(y)$ denoting the joint and marginal distributions of $X$ and $Y$, respectively. The \emph{channel} $C$ is characterised by the conditional probability distribution $p(y|x)$ and is assumed fixed by the data sharing system and publicly known. For any prior $p(x)$, a joint $p(x,y)$ and a marginal distribution $p(y)$ are induced by $C$.

Papers such as \cite{Smith09,issa2019operational} discuss some of the limitations of Shannon mutual information in properly quantifying or differentiating various adversarial threats to privacy. Sibson and Arimoto  mutual information of order $\alpha$ have recently been proposed to measure a spectrum of generalised threats to privacy \cite{2019TunMsurInfLeak_PUT}. It also turns out that the ``nucleus" of Shannon mutual information, namely the information density  variable: $i(x,y)\Defs \log\frac{p(x,y)}{p(x)p(y)}$ can represent a much stronger notion of privacy leakage than its average. We call $\ell(x,y)\Defs \frac{p(x,y)}{p(x)p(y)}$ the lift variable, the exponential of information density.~\footnote{Therefore, we may refer to information density as log-lift.} Information density and lift have been studied, albeit under various names, in works such as \cite{2012_privacy_statisticalinference,salamatian2020privacy,2021Contextaware,hsu2019information,sadeghi:ding:itw:2020}.

Differential privacy (DP) is a well-known privacy framework \cite{Dwork06} describing a worst-case attack scenario, in which an attacker with knowledge of all individuals but one in a dataset learns little information about the unknown individual upon an information release. Specifically, for a given $\epsilon$, any $E \subset \caly$ and any two neighbouring datasets $x,x'$ differing in one individual and any observation $Y \in \caly$, DP requires 
\begin{equation}\label{eqn:dp}
    p(Y\in E|x) \leq e^{\epsilon}p(Y\in E|x') .
\end{equation}
This definition is sometimes called the central model for differential privacy, referring to its dependence on a trusted centralised data curator. Alternatively, in the local model for differential privacy~\cite{LDP1,LDP2013MiniMax}, each individual first applies noise to their data before sending it to an \emph{untrusted} data curator, who releases some aggregate information about the (noisy) data. Local DP is more stringent than DP as it requires the inequation above be satisfied for all $x,x' \in \calx$. In other words, local $\epsilon$-DP implies central $\epsilon$-DP, but not vice versa.

Finally, quantitative information flow (QIF) is an information-theoretic framework for measuring leakage from (probabilistic) systems modelled as information-theoretic channels.~\footnote{The early work introducing Quantitative Information Flow was based on Shannon entropy to measure flows of information~\cite{qif2005}. The QIF used here refers to later work initiated by Smith \cite{Smith09}.} In QIF, leakages are derived from a Bayesian adversary model incorporating a gain function describing the adversary's goal, and the adversary's prior knowledge.  The resulting leakage, computed as the \emph{expected} gain in the adversary's knowledge, thus provides a meaningful, interpretable measure of the security risk to a system from that adversary. For example, the Bayes vulnerability measure, widely used in QIF, corresponds to the maximum probability that an adversary can guess the secret in one try; the Bayes leakage measure is the difference between the adversary's expected probability of guessing the secret \emph{after} making an observation from the system versus their probability \emph{before} making an observation (i.e., using only their prior knowledge).

\subsection{Explaining epsilon: a motivating example}\label{sec:motivating_example}

An important development in both the information theory and security communities has been an interest in operationally relevant leakage measures to provide information about the relative security of different probabilistic data sharing systems. In this context, \emph{operational} measures are ones which provide interpretable guarantees in terms of an adversary's gain or loss upon interacting with the system. The value produced by such a leakage measure therefore has a correspondence with some measurable information gain to the adversary, justifying its use as a security measure. For both of the above communities, the $\epsilon$ of differential privacy still lacks a robust operational meaning, despite the privacy community's explanation in terms of the differential privacy attacker who knows all-but-one secret in the dataset. The following example illustrates this more clearly.
 
 Consider a local differential privacy scenario in which a user wishes to obfuscate their own sensitive data before releasing it to an untrusted data curator. The user is given a choice of two mechanisms to use. The scenario is depicted by the two mechanisms below, in which, for example, the user is answering a question in a survey, and the values \{y, m, n\}  correspond to the answers `yes', `maybe' and `no', respectively. Note that we describe our mechanisms as `channels', in which the row labels correspond to secrets and the column labels correspond to observations. The value in row-$i$ column-$j$ represents the conditional probability $P(j | i)$; that is the probability of observing output $j$ given input $i$. In the example below, in the channel $G$ we have that the probability of observing $y$ given an input $n$ is $\nicefrac{1}{6}$, given by the value in the bottom left of the channel matrix.

\[
   \begin{array}{|c|ccc|}
            \hline
            G & y & m & n \\
            \hline
            y & \nicefrac{2}{3} & \nicefrac{1}{6} & \nicefrac{1}{6} \\
            m & \nicefrac{1}{3} & \nicefrac{1}{3} & \nicefrac{1}{3} \\
            n & \nicefrac{1}{6} & \nicefrac{1}{6} & \nicefrac{2}{3} \\
            \hline
        \end{array}
        \quad
        \begin{array}{|c|ccc|}
            \hline
            R & y & m & n \\ 
            \hline
            y & \nicefrac{3}{5} & \nicefrac{1}{5} & \nicefrac{1}{5} \\
            m & \nicefrac{1}{5} & \nicefrac{3}{5} & \nicefrac{1}{5} \\
            n & \nicefrac{1}{5} & \nicefrac{1}{5} & \nicefrac{3}{5} \\
            \hline
        \end{array}
\]

From the perspective of differential privacy, the choice of the user's noisy mechanism should be determined by the $\epsilon$ parameter described by the DP equation (\ref{eqn:dp}); in the DP literature, the smaller the $\epsilon$, the more privacy is said to be guaranteed to the user.

In the mechanisms described above, notice that the mechanism $G$ satisfies $\log(4)$-DP and the mechanism $R$ satisfies $\log(3)$-DP.~\footnote{The mechanisms $G$ and $R$ are in fact instances of the well-known Geometric and Randomised-Response mechanisms from DP, respectively.} (In fact this corresponds to the maximum ratio between any two values within a column). We would therefore expect that $R$ is more private and so leaks less information to an adversary than does $G$.

We now ask: for each of these mechanisms, what would be the expected gain to an attacker who is trying to guess the secret in one try? We can model this attack by assuming a uniform prior for the adversary (say), and applying Bayes rule to compute the adversary's posterior knowledge after making an observation from each mechanism.~\footnote{This is the model used in the QIF framework and is described in more detail in \Sec{sec:gleakage}.} The posterior probability of success for the attacker is given by the maximum probability in each posterior, averaged by the marginal on that posterior~\footnote{In QIF this is called the posterior Bayes vulnerability.}. The resulting computation  yields that the leakage of the system computed using the adversary's expected posterior gain from $G$ is $1.67$ whereas the equivalent leakage from $R$ is $1.80$ (see \App{app:example1} for complete details). So, in fact, our attacker learns \emph{more} from $R$ than she does from $G$.

But since $\epsilon$ represents a ``max-case'' measure and the adversarial scenario above uses the \emph{expected} gain of the attacker, i.e., an ``average-case'' measure, we might think that the above anomaly occurs because the measures we are using are incompatible. Let us consider now a max-case scenario: an attacker whose advantage is measured by the \emph{maximum} probability of guessing the secret in one try regardless of which output was observed. In this scenario, the leakage to the attacker is given by the maximum leakage of any of the posteriors of the channel, independent of the marginal probability on that posterior. In this instance, we find that, again under a uniform prior, the max-case leakage computed using the posterior max-case vulnerability for $G$ is $1.71$ whereas for $R$ it is $1.80$ (see \App{app:example1} for full details). And so $R$ also leaks more information to this max-case adversary than does $G$, even though its $\epsilon$ suggests that $R$ should be safer than $G$.

These examples motivate the question: what does $\epsilon$ tell us about other kinds of attacking threats apart from the one modelled in the DP literature?~\footnote{The DP literature describes $\epsilon$ as a measure of indistinguishability against an attacker who knows every secret in the dataset except for one.} If $\epsilon$ is indeed a \emph{robust} measure for privacy, it should be useful for making judgments about many different kinds of privacy threats. One way of assessing robustness from the QIF perspective would be to ask: does $\epsilon$ correspond to a leakage bound for some class of (Bayesian) adversaries? If so, this would give $\epsilon$ a robust operational meaning, allowing its guarantees to be explained in terms of a large number of attacking threats relevant to the security community. For the privacy community, this may assist in providing some \emph{explainability} for $\epsilon$ in terms of its measurement of leakage. This could be used to guide privacy practitioners in determining a reasonable value for $\epsilon$, which is a problem that has been identified by, for example, the implementers of differential privacy for the US Census Bureau~\cite{garfinkel2018issues}. 
 
In this paper, we bring together the information-theoretic, differential privacy and QIF frameworks to show how the measures of lift from information theory and $g$-leakage from QIF can be used to bring a robust operational meaning -- an explanation in terms of a broad class of adversarial attacks -- to the parameter $\epsilon$ in (local) differential privacy.

\subsection{Key open questions and contributions of this paper}

Despite a large body of existing work, a number of questions remain open at the intersection of these frameworks.

While $\epsilon$ in both the local and central models of DP is agnostic to the adversarial gain function and prior knowledge, it is not clear whether $\epsilon$ has a robust interpretation in the QIF sense (i.e., in terms of protection against a class of adversaries as highlighted in the example earlier). Such a robust operational interpretation is missing from the DP literature.

In addition, it is still not clear whether the measure of log-lift (in this paper \emph{lift}) has an operational significance as a measure of privacy leakage or is capable of robustly quantifying maximum privacy leakage in strong adversarial threat models via the QIF framework. 

Last, the current theory of max-case leakage in QIF leaves open the question of how best to model the max-case adversarial threats or how to characterise max-case \emph{capacity} over all priors and a wide class of adversarial threats. 

The following contributions of this paper address these questions.

\noindent \textbf{Main contributions:}

\begin{enumerate}
\item We propose a new set of measures: max-case $g$-leakages (\Def{def:max_leak}), which are a subset of the set of general max-case measures for QIF, which have been extensively studied in \cite{Alvim20:Book}. The max-case $g$-leakages have a clear operational significance: they correspond to adversaries whose goal is to maximise their gain over any output. Note that $g$-leakages model information-theoretic Bayesian adversaries and therefore represent a very general threat model.
\item We prove the first robust and operationally meaningful interpretation of $\epsilon$ in local DP via the proposed max-case $g$-leakage framework. Specifically, we show that $e^\epsilon$ is exactly the max-case $g$-leakage capacity, which measures the maximum leakage of a system wrt\, any Bayesian attack modelled using either average-case or max-case $g$-leakage measures, and is robustly independent of adversarial gain functions and priors. 
\item We introduce the \emph{lift capacity} (\Def{def:lift_capacity}), which quantifies lift over all priors, and show that it is equivalent to $e^\epsilon$ in the local DP framework. This provides the missing connection between lift and $\epsilon$ which has been a subject of interest in the information theory literature.
\end{enumerate}

We remark that although one conclusion we draw -- that the $\epsilon$ of the DP framework is robustly independent of the adversary -- may seem expected or unsurprising, the method developed in this paper of obtaining the result via the QIF framework is non-trivial.

\subsection{Detail of technical contributions}

Technically speaking, the first contribution of this work is establishing a link between lift in information-theoretic privacy and a notion of max-case $g$-leakage, which we introduce in this paper via the QIF framework. Specifically, in \Def{def:max_leak} we introduce a max-case multiplicative $g$-leakage, denoted by $\MaxLeak_g(\pi, C)$, using the standard notion of $g$-vulnerability in QIF framework. We then show via \Thm{thm:lift_bounds} that lift upper bounds  $\MaxLeak_g(\pi, C)$ with respect to any gain function $g$ for any given prior $\pi$. In \Thm{thm:lift_as_max}, we show that lift 
 is indeed realisable
 as a max-case $g$-leakage
 for an adversary who chooses a suitable prior-dependent gain function.  
 
 In addition,  we establish an important result, linking all three information-theoretic privacy, local differential privacy and QIF frameworks in an operationally robust manner. Specifically in the information-theoretic privacy framework, we define in \Def{def:lift_capacity} the supremum of lift over all priors for a channel $C$ as \emph{lift capacity} and denote it by  $\MaxLift(C)$.  We then show in \Thm{thm:equiv} and Cor~\ref{cor:epsilon} that lift capacity is equivalent to $\epsilon$ in the local DP framework. Combining this result with \Thm{thm:lift_as_max} and \Thm{thm:lift_bounds} we conclude that lift capacity, aka $e^\epsilon$, is also equal to the max-case $g$-leakage capacity in the QIF framework. This gives lift a robust operational interpretation in terms of strong max-case adversarial threats and explains $\epsilon$ in local DP as a capacity measure from the lens of information theory and QIF. 
 
Finally, we address a question raised in \cite{issa2019operational} as to whether the $g$-leakage framework is sufficient to describe adversarial scenarios in which the attacker is trying to guess a randomised function of the secret. In fact the Dalenius leakage described in \cite[Ch 10]{Alvim20:Book} already addresses this in the case of average-case $g$-leakage, demonstrating that $g$-leakage is unchanged when considering randomised functions of the secret. We extend these results to the cases of lift and lift capacity, showing that -- as with the Bayes capacity -- both measurements are unaffected under such adversarial scenarios.  

 Table \ref{tab:orders} visualises existing results in the literature and our contributions. As seen from the Table, this paper bridges between several results in the information-theoretic privacy, differential privacy and QIF literature and establishes new relations between lift and other measures of privacy. Our new results, together with our consolidated summary of existing results, depicts a fuller picture on deep connections between the  information-theoretic, differential privacy and QIF frameworks.

 \begin{table}[!ht]
\center
\colorlet{highlight}{magenta}
\colorlet{lowlight}{darkgray}
\renewcommand{\arraystretch}{1.5}
\setlength{\tabcolsep}{0.6em}
\begin{tabular}{ c c c c c c}
  \hline
   \textcolor{lowlight}{$\Leak^{\times}_g(\pi, C)$} &  \textcolor{highlight}{$\underset{\textrm{\Lem{lem:leak2}}}{\leq}$} &  \textcolor{highlight}{$\MaxLeak_g(\pi, C)$} &  \textcolor{highlight}{$\underset{\textrm{\Thm{thm:lift_bounds}}}{\leq}$} &  \textcolor{lowlight}{$\Lift(\pi, C)$} &\\
    \textcolor{lowlight}{\rotatebox[origin=c]{270}{$\leq$}} &  
    		&  \makecell{\textcolor{highlight}{\rotatebox[origin=c]{270}{$\leq$}} \\[-0.3em] \textcolor{highlight}{{Cor~\ref{cor:mlift_bounds}}}} & & \makecell{\textcolor{highlight}{\rotatebox[origin=c]{270}{$\leq$}} 
    		\\[-0.3em] \textcolor{highlight}{{\Def{def:lift_capacity}}}} & \\
   \textcolor{lowlight}{$\MultBayes(\Dist, C)$} &  \textcolor{highlight}{$\underset{\textrm{\Lem{bayeslem}}}{\leq}$} & \textcolor{highlight}{$\MaxLift(C)$}  &  \textcolor{lowlight}{$=$}  & \textcolor{highlight}{$\MaxLift(C)$} & \textcolor{highlight}{ $\underset{\textrm{ Cor~\ref{cor:epsilon}}}{=}$}$e^\epsilon$   \\[0.5em]
   \hline
 \end{tabular}
 \\[0.5em]
 \caption{Relationship between leakages (top row) and capacities (bottom row). The \textcolor{highlight}{coloured} text indicates new contributions of this paper.}\label{tab:orders}
\end{table}

\subsection{Operational measures and their significance in information theory and security}

A key motivating example for the importance of operational security measures came from Geoffrey Smith~\cite{Smith09} who observed that traditional information-theoretic measures such as mutual information can underestimate the threat to a system. This work led to the study of $g$-leakages, which directly model attacking threats and provide leakage measures which can be interpreted in terms of a concrete information gain to an attacker. The relevance of such leakage measures to the security community is that they are explainable in terms of explicit adversarial threats, and therefore give meaningful security assessments. The study of $g$-leakages now sits under the QIF umbrella~\cite{Alvim:2012aa,Alvim20:Book}.

In parallel to the above, recent work from the information theory community has identified the importance of operational measures for information leakage~\cite{issa2019operational}. Of particular interest is the problem of providing an operational interpretation to privacy measures such as $\epsilon$ (in differential privacy) and log-lift (from information theory). Key work in this area by Issa et al.~\cite{issa2019operational} has independently discovered several results overlapping with QIF, which will be outlined later in this paper. Importantly, both communities use an information-theoretic and decision-theoretic (Bayesian) approach to modelling adversarial threats. Therefore, although in this paper we use the notation and framework of quantitative information flow, the results will be relevant and interpretable for the information theory community as well.

\section{Information-Theoretic Foundations for Privacy}\label{sec2}

\subsection{The channel model for quantitative information flow}

We adopt notation from QIF~\cite{Alvim20:Book}, a mathematical framework for studying and quantifying information leaks with respect to adversarial attack scenarios.

A probabilistic channel $C$ maps inputs (secrets) $x \in \calx$ to observations $y \in \caly$ according to a distribution $\Dist{\caly}$. In the discrete case, such channels are $\calx{\times}\caly$ matrices $C$ whose row-$x$, column-$y$ element $\CondYX$ is the probability that input $x$ produces observation $y$. The $x$-th row $C_{x,-}$ is thus a discrete distribution in $\Dist{\caly}$. We write $\calx \rightarrow \Dist{\caly}$ for the type of the channel $C$.

We can use Bayes rule to model an adversary who uses their observations from a channel to (optimally) update their knowledge about the secrets $\calx$. 
Given a prior distribution $\pi : \Dist{\calx}$ (representing an adversary's prior knowledge) and channel $C$, we can compute a joint distribution $J: \Dist(\calx{\times}\caly)$ where $J_{x,y} = \px \CondYX$. Marginalising down columns yields the $y$-marginals $\py = \sum_x \px \CondYX$ each having a posterior over $\calx$ corresponding to the posterior probabilities $P_{X|y}(x)$, computed as $\nicefrac{J_{x,y}}{\py}$ (when $\py$ is non-zero). We denote by $\delta^y$ the posterior distribution $P(X|y)$ corresponding to the observation $y$. The set of posterior distributions and the corresponding marginals can be used to compute the adversary's posterior knowledge after making an observation from the channel.

\subsection{Local differential privacy and lift}\label{sec:lip}

Local differential privacy (LDP), as applied by an individual, can be defined as a property of a channel $C:\calx \rightarrow \Dist{\caly}$ taking data $\calx$ to noisy outputs $\caly$.

\begin{definition}\label{def:ldp}
We say that channel $C$ satisfies $\epsilon$-LDP if 
\[
	C_{x,y} ~\leq~ e^\epsilon C_{x', y} \qquad \forall x, x' \in \calx, y \in \caly~.
\]
\end{definition}

\noindent A central quantity in this paper, which we call lift, was defined in \cite{issa2019operational} as:
\begin{definition}\label{def:lift}
Given a channel $C$ and prior $\pi: \Dist{\calx}$, lift is defined as 
\begin{equation}\label{lift}
	\Lift(\pi, C) \Wide\Defs \max_{\substack{x\in \calx, y \in \caly:\\ J_{x,y} >0}}\,\, \frac{\CondYX}{\py}~.
\end{equation}
\end{definition}

\noindent \textbf{Remark:} Two alternative expressions for the lift are 
\begin{align}\label{altlift}
	\Lift(\pi, C) \Wide\Defs & \max_{\substack{x\in \calx, y \in \caly:\\ J_{x,y} >0}}\,\, \frac{\delta^y_x}{\pi_x}~,\\
 \Lift(\pi, C) \Wide\Defs & \max_{\substack{x\in \calx, y \in \caly: \\J_{x,y} >0}}\,\, \frac{J_{x,y}}{\px \py}~.\,\label{joint_lift}
\end{align}
 
Notably, the intuition behind lift as expressed in Eqn.~\eqref{altlift} is that it measures the adversary's change in knowledge, through (multiplicative) comparison of her prior and posterior beliefs for each secret and observation. The observation providing the biggest ``knowledge gap'' or \emph{lift} thereby produces the most leakage.

Note that the argument of maximisation $\frac{J_{x,y}}{\px \py}$ in Eqn.~\eqref{joint_lift} is indeed 
\[
\ell(x,y) =\frac{p(x,y)}{p(x)p(y)}~,
\]
in plain probability notation as described in \Sec{s1827} and its logarithm is known as the information density. Here, we provide a brief account of some works which have used such quantities in defining privacy measures. 

In \cite{2021Contextaware}, a mechanism is said to provide $\epsilon$-local information privacy (LIP) if
\begin{align}\label{lift_UB_LB}
e^{-\epsilon}\leq \frac{\delta^y_x}{\pi_x}\leq e^{\epsilon}, \qquad \forall x \in \calx, y \in \caly.
\end{align}
 A similar definition was earlier studied in \cite{2012_privacy_statisticalinference} for sensitive features of datasets. In \cite{hsu2019information,sadeghi:ding:itw:2020}, $\epsilon$-lift was said to be satisfied if the logarithm of the above inequalities held true for a sensitive variable $S$ and useful variable $X$ according to the Markov chain $S\to X\to Y$. %

An important distinction between our definition of lift and the above works is that they imposed an additional lower bound, $e^{-\epsilon}$, on the ratio of the posterior to prior beliefs.  Whereas in this work, we are only concerned with the largest lift ratio (i.e., the upper bound), which we simply call lift (removing the logarithm). Our definition captures the notion of maximum realisable leakage \cite{issa2019operational}, which is proved in \cite[Theorem 13]{issa2019operational} to be equal to the lift as stated in our Definition \ref{def:lift}. This also coincides with the notion of almost-sure pointwise maximal leakage in \cite[Definition 4]{saeidian2022pointwise}.

In \cite{2021Contextaware,2012_privacy_statisticalinference,hsu2019information,salamatian2020privacy}, it is proven that  \Eqn{lift_UB_LB} implies $2\epsilon$-LDP. This bound can be improved through optimisation as shown in \cite{2021Contextaware}. However, we highlight that \Eqn{lift_UB_LB} and the aforementioned relations to LDP depend on the prior $\pi$. In \Sec{sec:lift_capacity}, we will establish robust results strongly and directly linking a prior-independent notion of \emph{lift capacity} to $\epsilon$-LDP. The  reverse direction, linking $\epsilon$-LDP to \Eqn{lift_UB_LB} is already strong: it has been shown that in works such as \cite{2012_privacy_statisticalinference,salamatian2020privacy,2021Contextaware} that $\epsilon$-LDP implies \Eqn{lift_UB_LB}.

\begin{example}[Running Example]\label{running_example} 

We will use the following example throughout to show how to compute our various measures. Consider the following channel $C$:
\[
\begin{array}{|c|cc|}
		\hline
		C & b & g  \\
		\hline
		b & \nicefrac{3}{4} & \nicefrac{1}{4} \\
		g & \nicefrac{1}{4} & \nicefrac{3}{4} \\
		bg & \nicefrac{19}{20} & \nicefrac{1}{20} \\
		\hline
	\end{array}
\]
\noindent The rows represent the (secret) eye colour of an individual (blue, green or blue-green) and the column labels represent the eye colour that the individual reports (blue or green). The channel $C$ then tells us, for example, that with probability $\nicefrac{19}{20}$ the individual will report eye colour blue when they in fact have eye colour blue-green; however they will report eye colour blue with probability $\nicefrac{1}{4}$ when they have eye colour green.

\noindent Let us imagine that our adversary (trying to guess the secret eye colour) has some prior knowledge given by $\pi = (\nicefrac{1}{4}, \nicefrac{1}{2}, \nicefrac{1}{4})$. We then compute the adversary's posterior knowledge given the prior and the channel as the set of posterior distributions: 
\[
	\begin{array}{|c|c|}
	\hline
	 \delta^b & \delta^g \\
	\hline
	\nicefrac{15}{44} & \nicefrac{5}{36} \\
	\nicefrac{5}{22} & \nicefrac{5}{6} \\
	\nicefrac{19}{44} & \nicefrac{1}{36} \\
	\hline
	\end{array}
\]
with corresponding marginals $p(b) =  \nicefrac{11}{20}$, $p(g) = \nicefrac{9}{20}$.

The $\epsilon$ for this channel (\Def{def:ldp}) is computed using the maximum ratio between elements in any column; we have

\[
	 \epsilon = \log (\frac{C_{g,g}}{C_{bg,g}}) = \log(15) = 2.71 ~.
\]

\noindent The lift, in contrast, requires a prior $\pi$ in order to be computed (\Def{def:lift}). Given the prior $\pi$ from above, we compute the lift as:
\[
	\Lift(\pi, C) = \max \{ \nicefrac{C_{bg, b}}{p(b)}, \nicefrac{C_{g,g}}{p(g)} \} = \nicefrac{19}{11}= 1.73 ~.
\]
\end{example}

\subsection{Operational scenarios and the $g$-leakage framework}\label{sec:gleakage}

An important development over the past decade in security has been the use of operationally relevant leakage measures to provide information about the relative security of different probabilistic systems. \emph{Operational} measures -- those which have a direct correspondence with an adversarial scenario -- are crucial to a proper understanding of the security properties of a system. Foundational work in this area by Geoffrey Smith~\cite{Smith09} has led to the study of $g$-leakage under the umbrella of QIF~\cite{Alvim:2012aa,Alvim20:Book}. %

In QIF, adversaries are modelled as Bayesian: they are equipped with a prior $\pi\In\Dist{\calx}$ over secrets $\calx$ and a gain function $g\In\calw{\times}\calx{\rightarrow}\RealNN$ (over \emph{actions} $\calw$ and secrets $\calx$), which models the gain to the adversary upon taking action $w \in \calw$ when the true value of the secret is $x \in \calx$.  Before observing an output from the system, the adversary's prior expected gain, which we call the expected \emph{prior vulnerability} of the secret, can be expressed as
\begin{equation}\label{vgprior}
		V_g(\pi) \Wide\Defs \max_{w \in \calw} \sum_{x \in \calx} \px g(w, x) ~.
\end{equation}

The adversary can use their knowledge of the system (modelled as a channel $C\In\calx{\rightarrow}\Dist{\caly}$) to maximise their expected gain after making an observation. This is called the expected \emph{posterior vulnerability} and is expressed as
\begin{align}\label{vgposterior}
	  V_g{\Hyper \pi C} \Wide\Defs & \sum_{y \in \caly} \max_{w \in \calw} \sum_{x \in \calx} \px \CondYX g(w, x) ~\\\Wide= &\sum_{y \in \caly} p(y)\max_{w \in \calw} \sum_{x \in \calx}  \delta_x^y~ g(w, x) ~\\ \Wide= &\sum_{y \in \caly} p(y) V_g(\delta^y)~,\label{vgposterior2}
\end{align}
where in the last equality, we have used the definition of vulnerability in Eqn.~\eqref{vgprior} for the posterior distribution, denoted $\delta^y$.
Finally, the difference between the prior and posterior vulnerabilities gives a measure of the leakage of the system to this adversary; the greater the difference, the better is the adversary able to use the transmitted information to infer the value of the secret. This can be computed multiplicatively as 
\begin{equation}\label{eqn:avg_leak}
\Leak^{\times}_g(\pi, C) \Wide\Defs \frac{V_g{\Hyper \pi C}}{V_g(\pi)} ~.
\end{equation}
The $g$-leakage framework models a wide variety of attack scenarios, including guessing the secret in $n$ tries, guessing a property of the secret or guessing a value close to the secret.~\footnote{In fact, it has been shown that any convex vulnerability function is expressible as a $g$-vulnerability~\cite{McIver:2014aa}.} Moreover, attached  to each leakage measure is an operational scenario given by the gain function and prior which describes a specific adversarial threat. 

An important notion in QIF is that of capacity, which corresponds to the maximum leakage of a system quantified over all priors, or all gain functions, or both. Capacities thus provide robust, interpretable leakage bounds; robust because they quantify over large classes of adversaries and therefore are ``robust'' to variations in adversarial assumptions; interpretable because the bounds provide security guarantees which can be explained in terms of the particulars of adversarial attacks. 

Of particular note is the Bayes capacity, defined as
\begin{align}\label{bayes}
	\MultBayes(\Dist, C) \Wide\Defs & \sup_\pi \sup_g \Leak^{\times}_g(\pi, C) \\ \Wide= & \sup_\pi \sum_y \max_x \px \CondYX ~\\
 \Wide= & \sum_y \max_x\, \CondYX 
 ,\label{bayes2}
\end{align}
where the first equality is proved in \cite{Alvim:2012aa} and the last equality is due to the fact that the capacity is realised under the uniform prior \cite[``Miracle Theorem'']{Alvim:2012aa}. The Bayes capacity has an important operational significance: it is a tight upper bound on the multiplicative leakage of any channel in the average sense, quantified over all priors and gain functions~\cite{Alvim20:Book}. In other words, there is no adversarial scenario, modelled using the expected gain of an adversary with any prior knowledge, for which the channel leakage exceeds the amount given by the Bayes capacity.
 This provides a robust leakage bound on the security risk to a system against \emph{any} average-case attacker. 

\subsection{Connecting $g$-leakage and lift}

We conclude this section by showing the connection between Bayes capacity and the information-theoretic measure lift: 
it turns out that lift is an upper bound on the Bayes capacity. The following lemma has been expressed in \cite[Corollary of Theorem 13]{issa2019operational}; our contribution here is an alternative proof of this result in a QIF formulation.

\begin{lemma}\label{bayeslem}
Given a channel $C\In\calx{\rightarrow}\Dist{\caly}$, for all priors $\pi\In\Dist{\calx}$ it holds that
\[
	\MultBayes(\Dist, C) ~\leq~ \Lift(\pi, C)~.
\]
\begin{proof}
We reason as follows:
\begin{Reason}
\Step{}{
	\Lift(\pi, C)
}
\StepR{$=$}{\Eqn{lift}}{
	\max_{x,y} \frac{\CondYX}{\py}
}
\StepR{$=$}{Since $\sum_{y'} p(y') = 1$}{
	\max_{x,y} \frac{\CondYX}{\py} \sum_{y'} p(y') 
}
\StepR{$=$}{max independent of $y'$}{
	\sum_{y'} p(y') \max_{x, y} \frac{\CondYX}{\py}
}
\WideStepR{$\geq$}{Taking max over each column instead of channel}{
	\sum_{y} \py \max_x  \frac{\CondYX}{\py}
}
\StepR{$=$}{Rearranging}{
	\sum_y \max_x \CondYX
}
\StepR{$=$}{\Eqn{bayes2}}{
	\MultBayes(\Dist, C)
}
\end{Reason}
\end{proof}
\end{lemma}

Further, we find that the bound is strict, in that lift can be strictly greater than Bayes capacity.  Recalling that Bayes capacity is an \emph{upper bound} on average-case $g$-leakage measures, this result then shows that lift cannot be represented as an average-case $g$-leakage measure.

To illustrate, from our running example (Ex.~\ref{running_example}), we can compute the Bayes capacity on $C$ as the sum of the column maxima of $C$, yielding
\[
	\MultBayes(\Dist, C) \Wide= \nicefrac{19}{20} + \nicefrac{3}{4} \Wide= 1.7~.
\] 
\noindent This means that the maximum leakage of $C$ wrt\, \emph{any} adversary measured using an average-case gain, and for any prior and gain function, is $1.7$. However, recall that we computed the lift under the prior $\pi = (\nicefrac{1}{4}, \nicefrac{1}{2}, \nicefrac{1}{4})$ as 1.73. This means that lift \emph{cannot} be expressible as an average-case $g$-leakage measure.

Next, we study the relationship between lift and \emph{max-case} measures of leakage. 

\section{On Max-case $g$-leakage Measures, Lift and Local DP}

The QIF measures we have reviewed thus far have been for average-case attacks; that is, they  model the \emph{expected} gain of an attacker. Local differential privacy and lift, however, are max-case notions; they describe the worst-case gain for an adversary after making an observation, regardless of its probability of occurring. Max-case measures provide an alternative perspective to average-case measures: the average-case can be seen as the perspective of a data curator whose interest is in protecting attacks against the entire dataset, whereas the max-case provides the perspective of an individual in the dataset whose concern is their particular data point. 

The theory of QIF has been extended to include max-case measures which quantify the gain to an adversary interested in only the worst-case leakage~\cite{Alvim20:Book,Chatzi:2019}. To model this, the max-case posterior vulnerability is defined as
\begin{equation}\label{maxcase}
	\VMax\Hyper{\pi}{C} \Wide\Defs \max_y V(\delta^y)~.
\end{equation}
This says that the max-case posterior vulnerability $\VMax$  is the maximum vulnerability $V$ computed over the posteriors, where $V$ is a vulnerability function defined on distributions. The theory of max-case vulnerability leaves open the question of how best to define $V$, aside from a technical result which says that $V$ should be \emph{quasi-convex} in order to satisfy basic axioms such as the data processing inequality and monotonicity~\cite{alvim2016axioms}. 

In this paper, we introduce the notion of \emph{max-case $g$-leakage} by choosing $V$ in \Eqn{maxcase} to be the standard $g$-vulnerability function $V_g$ defined in \Eqn{vgprior}. Note that $V_g$ is \emph{convex}~\cite{Alvim20:Book}, and therefore also quasi-convex, thus represents a valid choice of vulnerability function. This yields the following definition:

\begin{definition}[Max-case $g$-leakage]\label{def:max_leak}
Given a channel $C$, prior $\pi$ and gain function $g$, the max-case $g$-leakage of $C$ is defined as
\begin{align}\label{eqn:max_leak}
	\MaxLeak_g(\pi, C) \Wide\Defs \frac{\VgMax\Hyper{\pi}{C}}{V_g(\pi)} ~~=~~ \frac{\max_y V_g(\delta^y)}{V_g(\pi)}~,
\end{align}
where $V_g$ is the prior vulnerability function given in \Eqn{vgprior}.
\end{definition}

Comparing \Eqn{eqn:max_leak} to \Eqn{eqn:avg_leak} we see that the max-case $g$-leakage quantifies the difference between the prior vulnerability and the posterior vulnerability computed as the maximum vulnerability of any of the posteriors wrt\, the adversary's measure of gain and their prior knowledge.

We remark that the significance of our restriction of $V$ to $g$-vulnerability functions is that it carries with it an operational interpretation in terms of adversarial threats.~\footnote{That is, our $V$ describes an adversary in terms of a gain function $g$ and a prior $\pi$; other (strictly quasi-convex) vulnerability functions cannot be written in this way.} We also note that max-case $g$-leakage has yet to be studied in the literature. We will leave further discussion on this choice of $V$ to \Sec{sec:discussion}.

\subsection{Lift and max-case $g$-leakage}

Using \Def{def:max_leak} above, we now show that lift can be expressed as a max-case $g$-leakage.

\begin{theorem}[Lift as max-case $g$-leakage]\label{thm:lift_as_max}
For discrete domain $\mathcal{X}$, define the set of actions $\mathcal{W} = \mathcal{X}$ and the gain function $g_\pi{\In}\mathcal{W}{\times}{\mathcal{X}} \rightarrow \mathbb{R}_{\geq 0}$ as:
\begin{equation}\label{eqn:inv_gain}
     g_\pi(w, x) =  
        \begin{cases}
           \frac{1}{\pi_x}, & \textit{if } w = x \\
           0, & \textit{otherwise}
        \end{cases}
\end{equation}
\noindent where $\pi$ is the (full support) prior of the adversary using the gain function $g_\pi$. Then the max-case  $g$-leakage of any channel $C$ given a prior $\pi$ is equal to its lift. That is, $\MaxLeak_{g_\pi}(\pi, C) = \Lift(\pi, C)$.
\begin{proof}
For the gain function given above, the prior vulnerability $V_{g_\pi}(\pi)$ is:
\begin{align*}
    V_{g_\pi}(\pi)~~ &= ~~   \max_w \sum_x \pi_x g_\pi(w, x) \\
            ~~ &=~~  1 \qquad \textit{ for all } w    
\end{align*}
and the max-case posterior vulnerability is:
\begin{Reason}
\Step{}{    
	V_{g_\pi}^\mathrm{max}{\Hyper{\pi}{C}}
}
\StepR{$=$}{\Def{def:max_leak}}{
	\max_y V_{g_\pi}(\delta^y)
}
\StepR{$=$}{Expanding $\delta^y$}{
	 \max_y \max_w \sum_x \frac{\pi_x C_{x, y} g_\pi(w, x)}{p(y)}
}
\StepR{$=$}{Substituting \Eqn{eqn:inv_gain}}{
  	 \max_y \max_x \frac{C_{x, y}}{p(y)}
}
\StepR{$=$}{\Def{def:lift}}{
	\Lift(\pi, C)
}
\end{Reason}

\noindent Thus the max-case leakage is $\MaxLeak_{g_\pi}(\pi, C) = \frac{V_{g_\pi}^\mathrm{max}{\Hyper{\pi}{C}}}{V_{g_\pi}(\pi)} = \Lift(\pi, C)$.
\end{proof}
\end{theorem}

The significance of \Thm{thm:lift_as_max} is that it gives an operational meaning to lift in terms of the adversary realised by $V_{g_\pi}$.
To give some intuition, for some distribution $\sigma$, $V_{g_\pi}(\sigma)$ is given by $\max_x \frac{\sigma_x}{\pi_x}$; i.e., it measures the maximum ``surprise'' to the adversary, achieved on the secret $x$ for which the probability $\sigma_x$ most differs (multiplicatively) from the adversary's prior $\pi_x$.~\footnote{We remark that the gain function $g_\pi$ is also known as the distribution-reciprocal gain function~\cite[Ch 3]{Alvim20:Book}.}

Continuing with our running example (Ex. \ref{running_example}), we can compute the gain function $g_{\pi}$ for $\pi(b, g, bg) = (\nicefrac{1}{4}, \nicefrac{1}{2}, \nicefrac{1}{4})$ as $g_{\pi}(b, b) = 4$, $g_{\pi}(g, g) = 2$, $g_{\pi}(bg, bg) = 4$.

The prior vulnerability is
\[
	V_{g_{\pi}}(\pi) \Wide= \max_w \sum_x \pi_x g_{\pi}(w, x) \Wide= 1 ~.
\]

The posterior vulnerability is then given by 
\[
	V_{g_\pi}^\mathrm{max}{\Hyper{\pi}{C}} = \max_y V_{g_{\pi}}(\delta^y) 
\]
where 
\[
    V_{g_{\pi}}(\delta^b) = \max \{ \nicefrac{15}{11}, \nicefrac{5}{11}, \nicefrac{19}{11} \} = \nicefrac{19}{11}
 \]
 and
 \[
    V_{g_{\pi}}(\delta^g) = \max \{ \nicefrac{5}{9}, \nicefrac{5}{3}, \nicefrac{1}{9} \} = \nicefrac{5}{3}
 \]
And so $\MaxLeak_{g_\pi}(\pi, C) = \nicefrac{19}{11} = 1.73$ which is what we calculated for lift in Ex.~\ref{running_example}.

We find, however, that lift has a much stronger property: it is in fact an upper bound on \emph{any} max-case $g$-leakage measure (that is, using any gain function) with the minor restriction that the gain function must be non-negative.~\footnote{Note that this non-negativity restriction is also required in the case of Bayes capacity as an upper bound on average-case leakage.}

\begin{theorem}[Lift bounds max-case $g$-leakage]\label{thm:lift_bounds}
Define the max-case $g$-leakage $\MaxLeak_g$ as per \Def{def:max_leak}. Then for any non-negative gain function $g$, any prior $\pi$ and any channel $C$, it holds that $\MaxLeak_g(\pi, C) \leq \Lift(\pi, C)$.
\begin{proof}
We reason as follows:
\begin{Reason}
\Step{}{
    \max_y V_g(\delta^y)
}
\WideStepR{$=$}{\Eqn{vgprior}, expand $\delta^y$}{
    \max_y (\max_w \sum_x \dfrac{\pi_x C_{x, y} g(w, x)}{p(y)})
}
\WideStepR{$\leq$}{$g \geq 0$}{
    \max_y (\max_w \sum_x (\max_x \dfrac{C_{x, y}}{p(y)})  \pi_x g(w, x) )
}
\WideStepR{$=$}{$\sum_i c x_i = c \sum_i x_i$ and $\max_i c x_i = c \max_i x_i$}{
    \max_y (\max_x \dfrac{C_{x,y}}{p(y)}) \max_w (\sum_x \pi_x g(w, x) )
}
\WideStepR{$=$}{\Eqn{vgprior}, $\Lift$}{
    \Lift(\pi, C) V_g(\pi)
}
\end{Reason}
The result follows using  \Def{def:max_leak}.
\end{proof}
\end{theorem}

\Thm{thm:lift_bounds} tells us that lift is a surprisingly robust measure; for any prior, lift is a measure of the maximum $g$-leakage of a channel wrt\ any (non-negative) gain function. Moreover, this upper bound holds regardless of whether we consider average-case leakage or max-case leakage: the max-case we proved above (\Thm{thm:lift_bounds}); the average-case follows from the fact that lift upper bounds Bayes capacity (\Lem{bayeslem}) and Bayes capacity is, by definition, an upper bound on any average-case $g$-leakage, for any prior and any (non-negative) gain function.

To illustrate, consider again the running example (Ex.~\ref{running_example}). For the gain function we will choose the \emph{gid} function which models the adversary who wishes to guess the secret in one try. This is defined:
\begin{equation*}
     gid(w, x) =  
        \begin{cases}
           1 & \textit{if } w = x, \\
           0 & \textit{otherwise}
        \end{cases}
\end{equation*}
Then the max-case $gid$-leakage given the prior $\pi = ( \nicefrac{1}{4}, \nicefrac{1}{2}, \nicefrac{1}{4} )$ is given by:
\[
\MaxLeak_{gid}(\pi, C) = \frac{V^{\MAX}_{gid}{\Hyper{\pi}{C}}}{V_{gid}(\pi)} = \frac{5}{6} / \frac{1}{2} = 1.67~.
\]
And the lift we computed earlier as $1.73$, which is larger than the $gid$-leakage, as we expected.
\Thm{thm:lift_bounds} tells us that in fact for any gain function we choose and any prior, the lift will always be at least as large as the max-case $g$-leakage.

\subsection{Robustness results: lift capacity and epsilon}\label{sec:lift_capacity}

Typically, leakage measures (as defined e.g., in QIF) are not \emph{robust} -- that is, they depend on the specifics of the adversary (their prior and gain function), and therefore may not provide a reliable measure of the safety of a channel against different (Bayesian) adversaries having different prior knowledge and intent. A more robust way to characterise a channel's leakage properties is to measure its \emph{maximum} leakage~\footnote{Note the distinction between \emph{maximum leakage} and \emph{max-case leakage}: the former quantifies over all priors and gain functions; the latter uses a max-case posterior vulnerability measure.} -- that is, quantified over all priors, or over all gain functions, or both. Such quantities are termed \emph{capacities} (see \Sec{sec:gleakage}), and have been previously studied in the context of average-case leakage~\cite{Alvim:14:CSF}; the Bayes capacity (recall \Eqn{bayes}) is the most well-known and provides robust upper bounds on the average-case leakage of \emph{any} channel. 

There has, to date, been no study of max-case capacities. However,  \Thm{thm:lift_bounds} gives an immediate result in this direction -- it shows that lift can be seen an example of a max-case capacity, since it provides an upper bound on max-case leakages quantified over all gain functions. It follows also that, by quantifying over all priors as well, the resulting \emph{lift capacity} is an upper bound on all max-case $g$-leakages for all priors.

\begin{definition}\label{def:lift_capacity}
We define the lift capacity for a channel $C$ as
\[
	\MaxLift(C) \Defs \sup_{\substack{\pi\In\Dist{\calx}:\\ \pi_x > 0}} \Lift(\pi, C) \Wide{=} \sup_{\substack{\pi\In\Dist{\calx}:\\ \pi_x > 0}}\,\,\,\max_{\substack{x\in \calx, y \in \caly:\\ J_{x,y} >0}} \,\,\frac{\CondYX}{\py}~.
\]
\end{definition}

\noindent Immediate from this definition and \Thm{thm:lift_bounds} we get the following:

\begin{corollary}\label{cor:mlift_bounds}
For any channel $C$ and prior $\pi$ it holds that 
\[
    \MaxLeak_g(\pi, C) ~~\leq~~  \MaxLift(C) ~.
\]
\end{corollary}

We now show  that this so-called lift capacity also has an important relationship with the epsilon of local differential privacy.

\begin{theorem}\label{thm:equiv}
Let $C\In\calx{\rightarrow}\Dist{\caly}$ be a channel and $\epsilon \geq 0$. Then
\[
	\MaxLift(C) \leq e^\epsilon ~~ \textrm{iff}~~ \frac{C_{x,y}}{C_{x', y}} \leq e^\epsilon ~~ \textrm{for all }x, x' \in \calx, y \in \caly~.
\]
\begin{proof}
For the forward direction, we first note that $\inf_\pi \sum_x \pi_x C_{x,y}$ occurs when $\pi$ is a point prior on $\argmin_x C_{x,y}$.
However, recall from Definition~\ref{def:lift_capacity} that $\MaxLift(C)$ is a supremum over all \emph{full support} priors. 
Let $x^* = \argmin_{x'} C_{x',y}$ 
; now define a sequence of priors 
\[
\pi^n_{x^*} = 1{-}\frac{1}{n} ~~ \textit{and}~~
\pi^n_x = \frac{1}{n(|\calx|-1)} ~~ \textit{for } x\neq x^*~.
\]
We see that $\pi^n$ is full support, and also $\lim_{n\to \infty}\sum_x \pi^n_x \CondYX = \min_x \CondYX$.
Therefore
\begin{equation}\label{eq:point}
    \inf_{\pi:\pi_x > 0} \sum_x \pi_x C_{x,y} = \min_x C_{x,y} ~.
\end{equation}

We also note that we can rewrite $p(y)$ as $\sum_{x'} \pi_{x'} C_{x',y}$. Therefore we have:
\begin{Reason}
\Step{}{
	\sup_{\pi:\pi_x > 0} \Lift(\pi, C)
}
\StepR{$=$}{\Eqn{lift}}{
    \sup_{\pi:\pi_x > 0} \max_{x,y} \frac{\CondYX}{\py}
}
\WideStepR{$=$}{Substituting for $p(y)$ as noted above}{
   \sup_{\pi:\pi_x > 0} \max_{x,y} \frac{C_{x,y}}{\sum_{x'} \pi_{x'} C_{x',y}}
}
\StepR{$=$}{Rearranging}{
    \max_{x,y} \sup_{\pi:\pi_x > 0}  \frac{C_{x,y}}{\sum_{x'} \pi_{x'} C_{x',y}}
}
\StepR{$=$}{$C_{x,y}$ independent of $\pi$}{
   \max_{x,y} ( \frac{C_{x,y}}{\inf_{\pi:\pi_x > 0} (\sum_{x'} \pi_{x'} C_{x',y})})
}
\StepR{$=$}{From \Eqn{eq:point} above}{
    \max_{x,y} \frac{C_{x,y}}{\min_{x'} C_{x',y}}
}
\StepR{$\geq$}{$\min_x \CondYX \leq \CondYX$}{
    \frac{C_{x,y}}{C_{x',y}} ~~ \textrm{ for all } x, x', y
}
\end{Reason}
And so we have that $\MaxLift(C) \leq e^\epsilon$ implies $\frac{C_{x,y}}{C_{x',y}} \leq e^\epsilon$ for all $x, x', y$.
For the reverse direction, we refer to previous works in which this has been shown to hold \cite{2012_privacy_statisticalinference,salamatian2020privacy,2021Contextaware}. 

\end{proof}
\end{theorem}

\begin{corollary}\label{cor:epsilon}
The lift capacity coincides with the $\epsilon$ value of an LDP channel. That is,
	$\MaxLift(C) = e^\epsilon$.
\begin{proof}
From \Thm{thm:equiv} we deduce
that $\MaxLift(C) = \sup_{x,x',y} \frac{C_{x,y}}{C_{x',y}}$.
\end{proof}
\end{corollary}

\Thm{thm:equiv} and Cor~\ref{cor:epsilon} establish the equivalence between the lift capacity and the epsilon parameter of local differential privacy, and by \Thm{thm:lift_as_max} and \Thm{thm:lift_bounds} this gives that epsilon (or rather, $e^\epsilon$) is a \emph{tight} upper bound on the max-case $g$-leakage of any channel. That is, $e^\epsilon$ is the max-case $g$-leakage capacity, quantified over all priors and gain functions. This result connecting $\epsilon$ with the $g$-leakage framework provides the first robust, operational interpretation for local differential privacy in terms of Bayesian adversarial threats. 

We note that $\epsilon$ has previously been interpreted as a capacity under QIF~\cite{Chatzi:2019}. However, in that work the vulnerability function chosen was not expressible as a $g$-vulnerability, and therefore did not carry with it the operational interpretation attached to the $g$-leakage framework. This was Smith's original motivation for a $g$-leakage based model for secure information leakage measurements.

We also remark that the relationship between lift and local differential privacy has been established previously (eg. \cite{issa2019operational,andres2013geo}), but these results have not been tied back to the $g$-leakage framework, and did not establish the operational significance of $\epsilon$ in terms of max-case $g$-leakage measures.

\subsection{Interpreting the results connecting $g$-leakage, lift capacity and epsilon}

\Thm{thm:lift_bounds} above tells us that lift bounds max-case $g$-leakage. To give some intuition for what this means in practice, suppose that we are worried about the leakage of a system to some attacker, both wrt\, any single individual in our system (i.e., the max-case leakage to the attacker) \emph{and} wrt\, the system as a whole (i.e., the attacker's expected gain). Let's also assume that we 
can estimate the prior knowledge of the attacker, but we do not know exactly what the attacker is trying to achieve (i.e., what is her gain function). What \Thm{thm:lift_bounds} tells us is that without knowing anything about the attacker's goal, we can still compute an upper bound on the leakage of the system to this attacker (i.e., we can measure how much more the attacker can gain by having access to the system). This upper bound is given by the \emph{lift} - a quantity independent of the gain function - and this is an upper bound on both average-case and max-case leakage to Bayesian adversaries.

Now let's assume that we are worried about the same attacker, however this time we know nothing about her prior knowledge nor her specific goal. What \Cor{cor:epsilon} tells us is that we can still  compute an upper bound on the leakage of the system to this attacker, and in this case the upper bound is given by the lift capacity, or alternatively by  $e^\epsilon$ (computed in the local differential privacy sense). This upper bound is robust, in the sense that it makes no assumptions about the adversary's prior knowledge or goal; thus it is termed a \emph{capacity}, consistent with the QIF literature. The $\epsilon$ capacity differs from the existing Bayes capacity in QIF, in that Bayes capacity is an upper bound on adversaries whose leakage is computed in the average-case, whereas the $e^\epsilon$ upper bound is a capacity on adversaries computed using either average-case or max-case leakage measures. The significance of these results is that we can now justify $\epsilon$ as a robust security parameter with an operational interpretation in terms of a large class of Bayesian attackers.

Recall the motivating example presented in the introduction (\Sec{sec:motivating_example}) in which a user is given a choice of local differential privacy mechanisms $G$ and $R$ with $\epsilon$ values of $\log(4)$ and $\log(3)$, respectively. Remember that we found that $R$ had a larger leakage (wrt\, both an average-case and a max-case Bayesian attacker) than $G$, even though it has a smaller $\epsilon$. The results of \Thm{thm:lift_bounds} and \Cor{cor:epsilon} \emph{do not} tell us when one mechanism will be better than another against a particular attack; for this we would turn to the refinement orders of QIF (studied in \cite{Chatzi:2019}) which tell us when one mechanism is better than another against \emph{all} attacks; or alternatively we would compute the leakage of each mechanism wrt\, an individual attack of concern (as we did in the introduction).

What \Thm{thm:lift_bounds} and \Cor{cor:epsilon} \emph{do} tell us is that there cannot be any attack (max-case or average-case) which produces more leakage than what the lift tells us (if we know the attacker's prior), or what $e^\epsilon$ tells us (if we do not). In the case of $R$ and $G$, \Cor{cor:epsilon} tells us that there does not exist an attack (max-case or average-case, and regardless of the adversary's prior) that produces more leakage than $e^\epsilon = 3$ or $e^\epsilon = 4$, respectively. Moreover these bounds are tight: there exists an attack which \emph{does} induce an $e^\epsilon$ leakage (and, in fact, \Thm{thm:equiv} gives a construction for such an attacker). This means that we can conclude that $R$ is \emph{generally} safer than $G$, because its leakage never exceeds $3$, even though $G$ may be better than $R$ in some specific circumstances.

Notice that the leakages we computed in the example in \Sec{sec:motivating_example} for $G$ were $1.67$ and $1.71$, and for $R$ they were both $1.80$. These are indeed lower than the corresponding $e^\epsilon$ bounds of $4$ and $3$, respectively. Interestingly, if we compute the Bayes capacity for $G$ and $R$, we find that these are $1.67$ and $1.80$, respectively (because the Bayes capacity is realised on a uniform prior for exactly the adversarial scenario we computed). What this tells us is that if we are only concerned with attackers modelled in the average-case, then $G$ is in fact \emph{generally} better than $R$ (but may be worse for particular attackers). The leakage of $1.71$ that we measured for $G$ was for a max-case attacker; and in fact the max-case leakage is always at least as large as the average-case leakage for any prior and any gain function, as we will prove later (see \Lem{lem:leak2} in \Sec{sec:discussion}).


\subsection{Max-case Dalenius leakage}

Up until this point, we have been satisfied with computing leakage measures with respect to deterministic functions of the domain $\calx$ (described by a prior $\pi$). As pointed out in \cite{issa2019operational}, one may also be concerned about potentially randomised functions of $\calx$. The QIF theory of Dalenius leakage~\cite[Ch 10]{Alvim20:Book} accounts for such functions by modelling the leakage that a channel $C{\In}\calx{\rightarrow}\Dist{\caly}$ induces on a secret $\calz$ through an unknown correlation  $J{\In}\Dist{(\calz{\times}\calx)}$. Given such a correlation $J$, we can factorise it into a prior $\rho{\In}\Dist{\calz}$ and a channel $D{\In}\calz{\rightarrow}\Dist{\calx}$ so that $J_{z,x} = \rho_z {\times} D_{z, x}$. 

In \cite{Alvim20:Book} it is shown that the multiplicative Bayes capacity of the combined system $DC$ (writing $DC$ for matrix multiplication) is in fact bounded from above by the multiplicative Bayes capacity of $C$; and thus considering arbitrary  randomised functions of the secret $\calx$ is not necessary for quantifying the maximum multiplicative leakage of a channel.

Here we confirm that this property also holds for the max-case leakage capacity (given by $\MaxLift$), and also by lift itself. This means that the capacity results proven in this paper also hold for randomised functions of $\calx$, and therefore  there is no advantage in considering randomised functions of the secret space. 

The max-case $g$-leakage of $\calz$ caused by $J$ and $C$ can be written as $\MaxLeak_g(\rho, DC)$. We now have the following:


\begin{lemma}\label{thm:dalenius_max}
Let $C{\In}\calx{\rightarrow}\Dist{\caly}$ be a channel and let $J{\In}\Dist{({\calz}{\times}{\calx})}$ be a correlation that factorises as $\rho \triangleright D$. Then 
\[
	\Lift(\rho, DC) \Wide{\leq} \min\{ \Lift(\rho, D), \Lift(\pi, C) \} ~,
\]
where $\pi{\In}\Dist{\calx}$ and $\pi_x = \sum_z \rho_z D_{z, x}$.
\begin{proof}
Referring to \Thm{thm:lift_as_max}, write $g_\rho$ for the gain function that realises lift. i.e., $\Lift(\rho, C) = \MaxLeak_{g_\rho}(\rho, C)$.
The data processing inequality for max-case leakage 
gives that 
\[
	\MaxLeak_{g_\rho}(\rho, DC) \Wide{\leq} \MaxLeak_{g_\rho}(\rho, D) 
\] for any $\rho$. Thus, from \Thm{thm:lift_as_max} we get
\[
	\Lift(\rho, DC) \Wide{\leq} \Lift(\rho, D)~.
\]

\noindent We next have that:
\begin{Reason}
\Step{}{
	\Lift(\rho, DC)
}
\WideStepR{$=$}{\Def{def:lift}}{
	\max_{\substack{z\in \calz, y \in \caly:\\ J_{z,y} >0}} \,\,\frac{(DC)_{z,y}}{\py}
}
\WideStepR{$=$}{Rewriting}{
	\max_{\substack{z\in \calz, y \in \caly:\\ J_{z,y} >0}} \,\,\frac{\sum_x D_{z,x}C_{x,y}}{\sum_z \rho_z \sum_{x'} D_{z, x'} C_{x',y}}
}
\WideStepR{$\leq$}{Taking max of $C_{-,y}$ over $\calx$}{
	\max_{\substack{z\in \calz, y \in \caly:\\ J_{z,y} >0}} \,\,\frac{\sum_x D_{z,x} (\max_x C_{x,y})}{\sum_z \rho_z \sum_{x'} D_{z, x'} C_{x',y}}
}
\WideStepR{$=$}{Factorise; use $\sum_x D_{z, x} = 1$}{
	\max_{\substack{z\in \calz, y \in \caly:\\ J_{z,y} >0}} \,\,\frac{\max_x C_{x,y}}{\sum_z \rho_z \sum_{x'} D_{z, x'} C_{x',y}}
}
\WideStepR{$=$}{Factorise denominator}{
	\max_{\substack{z\in \calz, y \in \caly:\\ J_{z,y} >0}} \,\,\frac{\max_x C_{x,y}}{\sum_{x'} \sum_z (\rho_z D_{z, x'}) C_{x',y}}
}
\WideStepR{$=$}{Substituting $\sum_z \rho_z D_{z, x} = \pi_x$}{
	\max_{\substack{z\in \calz, y \in \caly:\\ J_{z,y} >0}} \,\,\frac{\max_x C_{x,y}}{\sum_{x'} \pi_{x'} C_{x',y}}
}
\WideStepR{$=$}{No dependence on $\calz$}{
	\max_{\substack{x\in \calx, y \in \caly:\\ J_{x,y} >0}} \,\,\frac{C_{x,y}}{\sum_{x'} \pi_{x'} C_{x',y}}
}
\WideStepR{$=$}{\Def{def:lift}}{
	\Lift(\pi, C)
}
\end{Reason}
And thus $\Lift(\rho, DC) \leq \min\{ \Lift(\rho, D), \Lift(\pi, C) \}$.
\end{proof}
\end{lemma}

As a corollary, we have that the lift capacity also respects Dalenius leakage, and therefore
the max-case $g$-leakage of secrets $\calz$ via a channel $C$ and arbitrary correlation $J$ is bounded from above by the lift capacity of $C$.

\begin{corollary}\label{cor:dalenius_bound}
For any channel $C{\In}\calx{\rightarrow}\Dist{\caly}$, non-negative gain function $g$ and correlation $J$ given by $J_{z,x} = \rho_z D_{z,x}$ we have that
\[
	\MaxLeak_g(\rho, DC) \Wide{\leq} \MaxLift(DC) \Wide{\leq} \MaxLift(C)~.
\]
\begin{proof}
By Cor~\ref{cor:mlift_bounds}, $\MaxLeak_g(\rho, DC) \leq \MaxLift(DC)$. 
From \Lem{thm:dalenius_max} we have $\Lift(\rho, DC) \leq \Lift(\pi, C)$ and thus we deduce
\begin{Reason}
\Step{}{
	\MaxLift(DC)
}
\StepR{$=$}{\Def{def:lift_capacity}}{
    \sup_{\substack{\rho\In\Dist{\calz}:\\ \rho_z > 0}}\,\,\,\Lift(\rho, DC)
}
\WideStepR{$\leq$}{\Lem{thm:dalenius_max}, substituting $\pi_x = \sum_z \rho_z D_{z,x}$}{
   \sup_{\substack{\rho\In\Dist{\calz}:\\ \rho_z > 0}}\,\,\,\Lift(\pi, C)
}
\StepR{$=$}{Independence from $\calz$}{
  \sup_{\substack{\pi\In\Dist{\calx}:\\ \pi_x > 0}}\,\,\,\Lift(\pi, C)
 }
 \StepR{$=$}{\Def{def:lift_capacity}}{
 	\MaxLift(C)
}
\end{Reason}
\end{proof}
\end{corollary}

\subsection{Interpreting the Dalenius leakage results}

To provide some intuition on how to interpret the above results, we return to our motivating example from the introduction, in which the two mechanisms $G$ and $R$ were introduced. Recall that these mechanisms were applied to survey results with answers `yes', `maybe' and `no'. Now let us assume that the adversary knows a correlation between survey answers and disease; perhaps users who answered `yes' or `maybe' are likely to have some serious illness, whereas users who answered `no' are unlikely to have one. We wish to know whether such a correlation will cause extra harm to the individual: can the adversary learn more information (via the correlation) than the $\epsilon$ parameter of $G$ or $R$ suggests?

What Cor~\ref{cor:dalenius_bound} tells us is that the max-case $g$-leakage of the entire system (including the correlation) is bounded above by $e^\epsilon$ computed from the mechanism ($G$ or $R$). This means that any (potentially public) correlations $D$ that the adversary may have access to cannot increase the amount of leakage caused by $G$ or $R$ -- the upper bound on leakage remains intact, where here we measure leakage using either lift or lift capacity ($e^\epsilon$). In other words, the designer of the system can focus on the $\epsilon$ parameter describing the mechanism, and so long as the leakage (represented by $e^\epsilon$) is small enough, then the system is protected against an adversary who knows any arbitrary correlation $D$. Note that we did not prove a Dalenius leakage result on max-case $g$-leakage, which we leave to future work.

We remark that this result appears to be similar to the notion in differential privacy that $\epsilon$ is independent of the prior knowledge of an attacker. A difference here is that the notion of \emph{prior knowledge} is typically represented as a distribution over secrets; in the Dalenius scenario described above, we are interested in arbitrary correlations between the secret and some other (potentially damaging) information, and the concern is what damaging information the attacker can learn as a result of the data release and the correlation. Dalenius leakage reassures us that that the damage caused by any arbitrary correlation is always bounded by the leakage of the original data release.

Interestingly, it may be the case that these Dalenius leakage results do not hold in the general case of (central) differential privacy; a counter-example appears in the work of \cite{wang2017privacy}. We leave further exploration of this idea to future work.

\section{Additional Results on Max-case Leakage and Technical Discussion}\label{sec:discussion}

In this section, we provide further technical details on the max-case leakage definition of \Def{def:max_leak}.

Firstly, the following result shows that the max-case leakage of a channel is at least as large as the average-case leakage. This result completes Table~\ref{tab:orders}.

\begin{lemma}\label{lem:leak2}
Given a channel $C$ prior $\pi$ and gain function $g$, it holds that
\[
   \Leak_g^\mult(\pi, C) ~~\leq~~ \MaxLeak_g(\pi, C)~.
\]
\begin{proof}
We reason as follows:
\begin{Reason}
\Step{}{
    V_g\Hyper{\pi}{C}
 }
 \StepR{$=$}{\Eqn{vgposterior2}}{
   \sum_y p(y) V_g(\delta^y)
}
\StepR{$\leq$}{Max over posteriors}{
   \sum_y p(y) \max_j V_g(\delta^j)
}
\StepR{$=$}{Factorising}{
   \max_j V_g(\delta^j) \sum_y p(y)
}
\StepR{$=$}{Simplify, \Def{def:max_leak}}{
   \VgMax\Hyper{\pi}{C}
}
\end{Reason}
Thus the corresponding leakages are ordered (since prior vulnerability is common).
\end{proof}
\end{lemma}

Next, we recall that in \Def{def:max_leak} we chose to model max-case leakage using the prior vulnerability function $V_g$, which models the adversary's \emph{expected gain} before interacting with a system (i.e., using only their prior knowledge). However, in the max-case setting it might seem preferable to choose a prior vulnerability function which models the adversary's \emph{max-case} gain; i.e., we could have defined:

\begin{equation}\label{eqn:optional_vg}
	\VgMax(\pi) \Wide\Defs \max_{w,x} \pi_x g(w, x) .
\end{equation}

That is, the adversary's prior gain is computed using the secret $x$ which maximises their gain. We now justify our original decision (choosing an average-case prior vulnerability) by demonstrating that these choices are, in fact, equivalent.

\begin{lemma}\label{lem:equivalence}
Let $C$ be a channel, $\pi{\In}\Dist{\cal{X}}$ be a prior and $g$ be a gain function. Then there exists a gain function $g^*$ such that $\VgMax(\pi) = V_{g^*}(\pi)$ and $\frac{\max_y \VgMax(\delta^y)}{\VgMax(\pi)} = \frac{\max_y V_{g^*}(\delta^y)}{V_{g^*}(\pi)}$.
\begin{proof}{(Sketch)}
Observe that $\VgMax$ (\Eqn{eqn:optional_vg}) is convex in $\pi$ and so it can be expressed as a convex $g$-vulnerability~\cite[Ch 11, Thm 11.5]{Alvim20:Book};
i.e., there exists a gain function $g^*$ such that $V_{g^*}(\pi) = \VgMax(\pi)$.
This also means that $\max_y \VgMax(\delta^y) = \max_{y} V_{g^*}(\delta^{y})$ and thus 
\[
    \frac{\max_y \VgMax(\delta^y)}{\VgMax(\pi)} ~=~ \frac{\max_y V_{g^*}(\delta^y)}{V_{g^*}(\pi)} .
\]
In Appendix~\ref{app:max-case} we show the construction of such a $g^*$.
\end{proof}
\end{lemma}

Finally, we recall the result of \cite{alvim2016axioms} which shows that, for reasonable axioms to hold under a max-case definition of leakage, then the prior vulnerability function should be quasi-convex (in the prior). However, as we have seen, both the average-case and max-case prior vulnerability functions are \emph{convex}. An open question in the community has been: are there any strictly quasi-convex functions which produce leakage measures of interest? Previous work~\cite{Chatzi:2019} showed that the $\epsilon$ of metric differential privacy ($d$-privacy) can be expressed as an additive capacity using a quasi-convex vulnerability function, suggesting that this was indeed one such example. In this paper we have resolved this question with a much stronger result, showing that $\epsilon$ in local DP can in fact be expressed using a \emph{convex} vulnerability function. This justifies our restriction of max-case leakage to $g$-leakage measures, but leaves open the question of the usefulness of max-case leakage defined over the full scope of quasi-convex vulnerabilities.

\section{Prior Work}

Perhaps not surprisingly and underpinned by the foundational theory of information, many linkages have already been established between information-theoretic, differential privacy and QIF  frameworks \cite{Chatzi:2019,cuff:2016}. Here, we only review results that are most pertinent to this work.

First, the logarithm of the Bayes capacity
is known as the Sibson mutual information of order $\alpha = \infty$ in information theory~\cite{sibson1969information} and was recently shown in \cite{issa2019operational} to measure the maximum leakage of adversaries wanting to guess arbitrary randomised functions of secret $X$. It turns out that this identity  is no coincidence and a recent work \cite{saeidian2022pointwise} proves that there is no advantage in generalising the class of adversaries from those who use deterministic gain functions to those who guess randomised functions. We remark that this result is also known in the QIF community via Dalenius leakage~\cite[Ch 10]{Alvim20:Book}. 

Second, it is not difficult to prove that an upper bound on the information density 
$i(x,y)$ bounds the Sibson mutual information of order infinity \cite{issa2019operational}, aka the logarithm of Bayes capacity in QIF language. Moreover, works such as \cite{2012_privacy_statisticalinference,salamatian2020privacy,2021Contextaware} link upper and lower bounds on lift to (local) differential privacy measures \cite{Dwork06,LDP1,LDP2013MiniMax} and vice versa.

There have been a number of works connecting DP and QIF, in particular via the study of \emph{max-case} leakage measures~\cite{Alvim20:Book,Chatzi:2019}. A general notion of max-case leakage under QIF has been extensively explored in \cite{Alvim20:Book}; there it was found that max-case measures are required to be quasi-convex in order to satisfy certain axioms. Our work differs from theirs in that we restrict our attention to the set of max-case $g$-leakages -- that is, derived from $g$-vulnerability functions -- which are a subset of the quasi-convex max-case vulnerabilities (in fact, we have shown that our max-case $g$-leakages are all \emph{convex}). To our knowledge our work is the first to explore max-case $g$-leakages and their capacities.
General max-case leakages were also explored in~\cite{Chatzi:2019} and their connection to differential privacy established via a leakage ordering. In particular, \cite{Chatzi:2019} found that whenever there is a max-case leakage order between two mechanisms (meaning one is \emph{always} safer than the other wrt\, \emph{every} max-case adversarial scenario), then there is a corresponding $\epsilon$ ordering between the two. While \cite{Chatzi:2019} explored $\epsilon$ as inducing a refinement order, finding that it is surprisingly weak (i.e., compared with the refinement orders of average-case and max-case found in QIF), in our work we explore $\epsilon$ as a capacity, finding that it does have a strong and meaningful  interpretation (in terms of max-case attackers). In addition,  \cite{Chatzi:2019} established that the $\epsilon$ of DP can be expressed as a max-case leakage capacity (via the QIF framework) derived from a specific quasi-convex vulnerability function. Such quasi-convex vulnerabilities, unfortunately, do not correspond to adversarial models in $g$-leakage, and therefore fall outside the desirable operational interpretability.  Our work finds that $\epsilon$ (actually $e^\epsilon$) is a capacity which can in fact be represented by a max-case $g$-vulnerability, which brings with it a broad operational meaning in terms of max-case attackers.

Other relationships between Bayesian models and differential privacy have also been previously studied~\cite{DBLP:conf/icalp/AlvimACP11,andres2013geo}, however none of these results give both a robust and an interpretable meaning to $\epsilon$. Andres et al.~\cite{andres2013geo} provided a Bayesian interpretation for metric differential privacy in terms of a single adversary and Alvim et al.~\cite{DBLP:conf/icalp/AlvimACP11} showed that $\epsilon$-differential privacy implies a bound on leakage but that the converse does not hold.

\section{Conclusion}

In this paper we have investigated the relationships between traditional information theory and the $g$-leakage framework of quantitative information flow. The connections are summarised in Table~\ref{tab:orders}. Overall we observe that the two notions converge wrt.\  robustness, namely via the capacity 
$\MaxLift(C)$, which we find is equivalent to $e^\epsilon$ of local differential privacy.

Significant also is that local differential privacy's $\epsilon$ parameter can now be understood through the lens of QIF. In particular, we see that it is also a measure of robustness in that it behaves as a capacity -- that is, independent of any particular prior. Moreover, it represents a tight upper bound on all max-case $g$-leakages. This is the first time that $\epsilon$ in LDP has been explained as a robust measure of information leakage in the QIF framework.

From the perspective of information theory,  lift is also explained  as a leakage measure, but interestingly we discovered that it  has ``capacity-like-properties'' (\Sec{sec:lift_capacity}).  Table~\ref{tab:orders} clarifies how these measures relate to the leakages and capacities well-known in QIF.

Differential privacy is often seen as a useful technique to protect the privacy of individuals' data, and has been used in several prominent applications including the 2020 US Census \cite{garfinkel2018issues}. As noted however, in spite of differential privacy's theoretical properties, there remain a number of challenges for its successful application. One such challenge is how to choose an appropriate level of $\epsilon$ relative to a particular scenario.  Whilst $\epsilon$ itself provides information about indistinguishability, it is difficult to reconcile it with the protection against other relevant attacks described here.

The work presented here is an important step towards a better understanding of how to choose $\epsilon$ where indistinguishability is not the only concern, but where there are differing adversarial assumptions. The results here give a clear theoretical account of $\epsilon$ and how to view it under differing adversarial conditions which can then be included in the determination of an appropriate threat model that is relevant to the scenario.

While the results of this paper were derived for local DP, they provide some insights about central DP. For instance, based on \Thm{thm:equiv}, Cor~\ref{cor:epsilon} and the fact that local DP implies central DP, we know that lift capacity upper bounds $e^\epsilon$ in central DP. However, more work is required to better understand what type of worst-case threat models in QIF framework have a one-to-one correspondence with $\epsilon$ in central DP. If such equivalence were to be established, it would settle the question of  the operational meaning of central DP in terms of which adversarial threats it can protect against.  There may also exist new connections or applications of lift and lift capacity in defining or bounding other privacy and leakage measures. We leave these interesting questions for future work.

\bibliographystyle{ieeetr}
\bibliography{loglift}


\appendices

\section{Motivating Example from the Introduction}\label{app:example1}

In the introduction we motivated our work with the example of the following two channels:

\[
   \begin{array}{|c|ccc|}
            \hline
            G & y & m & n \\
            \hline
            y & \nicefrac{2}{3} & \nicefrac{1}{6} & \nicefrac{1}{6} \\
            m & \nicefrac{1}{3} & \nicefrac{1}{3} & \nicefrac{1}{3} \\
            n & \nicefrac{1}{6} & \nicefrac{1}{6} & \nicefrac{2}{3} \\
            \hline
        \end{array}
        \quad
        \begin{array}{|c|ccc|}
            \hline
            R & y & m & n \\ 
            \hline
            y & \nicefrac{3}{5} & \nicefrac{1}{5} & \nicefrac{1}{5} \\
            m & \nicefrac{1}{5} & \nicefrac{3}{5} & \nicefrac{1}{5} \\
            n & \nicefrac{1}{5} & \nicefrac{1}{5} & \nicefrac{3}{5} \\
            \hline
        \end{array}
\]

Using a uniform prior $\pi = [ \nicefrac{1}{3}, \nicefrac{1}{3}, \nicefrac{1}{3} ]$, we can compute the set of posteriors using Bayes rule by multiplying out the channel by the prior and then normalising down columns. This yields the following posteriors and their corresponding marginals:

\[
  \begin{array}{r}
    \Hyper{\pi}{G} = \begin{array}{|c|c|c|}
            \hline
             \delta^y & \delta^m & \delta^n \\
            \hline
            \nicefrac{4}{7} & \nicefrac{1}{4} & \nicefrac{1}{7} \\
            \nicefrac{2}{7} & \nicefrac{1}{2} & \nicefrac{2}{7} \\
            \nicefrac{1}{7} & \nicefrac{1}{4} & \nicefrac{4}{7} \\
            \hline 
        \end{array} \\[25pt]
        p(y) = \begin{bmatrix}
         \nicefrac{7}{18} & \nicefrac{2}{9} & \nicefrac{7}{18}
          \end{bmatrix}
      \end{array}
        \quad
      \begin{array}{r}
        \Hyper{\pi}{R}  = \begin{array}{|c|c|c|}
            \hline
            \delta^y & \delta^m & \delta^n \\ 
            \hline
            \nicefrac{3}{5} & \nicefrac{1}{5} & \nicefrac{1}{5} \\
            \nicefrac{1}{5} & \nicefrac{3}{5} & \nicefrac{1}{5} \\
            \nicefrac{1}{5} & \nicefrac{1}{5} & \nicefrac{3}{5} \\
            \hline
        \end{array}\\[25pt]
        p(y) = \begin{bmatrix}
          \nicefrac{1}{3} & \nicefrac{1}{3} & \nicefrac{1}{3}
          \end{bmatrix}
      \end{array}
\]

We can now compute the expected posterior gain to an adversary who wishes to guess the secret in one try. This adversary can be modelled using the gain function \emph{gid} defined as:

\begin{equation}
     gid(w, x) =  
        \begin{cases}
           1 & \textit{if } w = x, \\
           0 & \textit{otherwise}
        \end{cases}
\end{equation}

We first compute the prior vulnerability for this adversary, which gives a measure of the leakage before observing any output from the channel. This is computed as: 

\begin{align*}
   V_{gid}(\pi) &= \max_w \sum_x \pi_x \textit{gid}(w, x)  \\
                      &= \max_x \pi_x \\
                      &= \nicefrac{1}{3}              
\end{align*}

The posterior vulnerability determines the expected gain to the adversary after making an observation, and therefore depends on the channel. The posterior \emph{gid}-vulnerability for $G$ is then:

\begin{align*}
	V_{gid}{\Hyper{\pi}{G}} &= \sum_y p(y) V_{gid}(\delta^y) \\
	    &= \nicefrac{7}{18}{\times}\nicefrac{4}{7} + \nicefrac{2}{9}{\times}\nicefrac{1}{2} + \nicefrac{7}{18}{\times}\nicefrac{4}{7} \\
	    &= \nicefrac{5}{9} \\
	    &= 0.56
\end{align*}

And for $R$ we have:

\begin{align*}
	V_{gid}{\Hyper{\pi}{R}} &= \sum_y p(y) V_{gid}(\delta^y) \\
	    &= \nicefrac{1}{3}{\times}\nicefrac{3}{5}{\times}3 \\
	    &= \nicefrac{3}{5} \\
	    &= 0.6
\end{align*}

The multiplicative leakage is given by the ratio of the posterior to the prior vulnerabilities. For $G$ we have:
\[
   \mathcal{L}^{\times}_{gid}(\pi, G) = \frac{V_{gid}{\Hyper{\pi}{G}}}{V_{gid}(\pi)} = \frac{15}{9} = 1.67
\]

And for $R$ we compute:
\[
   \mathcal{L}^{\times}_{gid}(\pi, R) = \frac{V_{gid}{\Hyper{\pi}{R}}}{V_{gid}(\pi)} = \frac{9}{5} = 1.80
\]

And thus $R$ leaks more than $G$ to this adversary.

Next we compare the max-case leakage using the same gain function \emph{gid} and the same uniform prior $\pi$.  Note that the prior max-case $gid$-vulnerability is the same ($\nicefrac{1}{3}$). For the posterior vulnerability for $G$ we compute:

\begin{align*}
	V^{\MAX}_{gid}{\Hyper{\pi}{G}} &= \max_y  V^{\MAX}_{gid}(\delta^y) \\
	    &= \max \{ \nicefrac{4}{7},  \nicefrac{1}{2}, \nicefrac{4}{7} \} \\
	    &= \nicefrac{4}{7} \\
	    &= 0.57
\end{align*}

And for $R$ we have:

\begin{align*}
	V^{\MAX}_{gid}{\Hyper{\pi}{R}} &= \max_y  V^{\MAX}_{gid}(\delta^y) \\
	    &= \nicefrac{3}{5} \\
	    &= 0.6
\end{align*}

And so the corresponding multiplicative leakage for $G$ is: 

\[
    \MaxLeak_{gid}(\pi, G) = \frac{V^{\MAX}_{gid}{\Hyper{\pi}{G}}}{V_{gid}(\pi)} = \frac{12}{7} = 1.71 
\]

And for $R$ it is:

\[
    \MaxLeak_{gid}(\pi, R) = \frac{V^{\MAX}_{gid}{\Hyper{\pi}{R}}}{V_{gid}(\pi)} = \frac{9}{5} = 1.80 
\]

And so again we find that $R$ leaks more than $G$ for this adversary, now modelled using a max-case vulnerability function.

\section{Max-case $g$-leakage using average-case $g$-vulnerability}\label{app:max-case}

In this section, we complete the proof of \Lem{lem:equivalence}, showing a gain function $g^*$ which produces the same average-case leakage as a max-case leakage defined using a $g$. %
That is, we will prove that for any gain function $g$ it is possible to construct a gain function $g^*$ such that $\VgMax(\pi) = V_{g^*}(\pi)$ where $V_{g^*}(\pi)$ is defined in the usual way as:
\begin{equation}\label{eqn:vgstar}
    V_{g^*}(\pi) \Wide\Defs \max_{w} \sum_x \pi_x g^*(w, x)~,
\end{equation}
and $\VgMax(\pi)$ is defined as 
\begin{equation}\label{eqn:vgmax_app}
   \VgMax(\pi) \Wide\Defs \max_{w, x} \pi_x g(w, x)~.
\end{equation}

Note that our construction assumes that the domain $\calx$ is discrete, although the proof does not rely on this assumption.

For $g{\In}\calw{\times}\calx \rightarrow \mathbb{R}_{\geq 0}$ we can define the set of actions ${\calw}^*$ such that for each $w \in \calw$ we have a set $\{w_x \in {\calw}^* : x \in \calx\}$, and a mapping $g^*{\In}{\calw}^*{\times}\calx \rightarrow \mathbb{R}_{\geq 0}$ satisfying 
\begin{equation}\label{eqn:gstar}
     g^*(w_{x_i}, x) =  
        \begin{cases}
           g(w, x), & \textit{if } x = x_i \\
           0, & \textit{otherwise.}
        \end{cases}
\end{equation}

\noindent This means that $\max_x g^*(w_{x_i}, x) = \max_x g(w, x)$.
And therefore: 

\begin{Reason}
\Step{}{
	\VgMax(\pi)
}
\StepR{$=$}{From \Eqn{eqn:vgmax_app}}{
    \max_{w \in \calw, x \in \calx} \pi_x g(w, x)
}
\StepR{$=$}{From \Eqn{eqn:gstar}}{
    \max_{w \in {\calw}^*, x \in \calx} \pi_x g^*(w, x)
}
\WideStepR{$=$}{Since $g^*(w, x) = 0$ everywhere else}{
    \max_{w \in {\calw}^*} \sum_x \pi_x g^*(w, x)
}
\StepR{$=$}{\Eqn{eqn:vgstar}}{
    V_{g^*}(\pi)
}
\end{Reason}

And this gives that $\MaxLeak_g(\pi, C) = \frac{\max_y \VgMax(\delta^y)}{\VgMax(\pi)} = \frac{\max_y V_{g^*}(\delta^y)}{V_{g^*}(\pi)}$.

\end{document}